\documentclass[doublecol]{epl2}
\usepackage{scalefnt}

\title{A nonextensive view of the stellar braking indices}
\shorttitle{A nonextensive view of the stellar braking indices}

\author{D. B. de Freitas$^{1}$\thanks{E-mail: \email{danielbrito@dfte.ufrn.br}}
\and F. J. Cavalcante$^{1}$
\and B. B. Soares$^{2}$
\and J. R. P. Silva$^{2}$}
\shortauthor{D. B. de Freitas \etal}

\institute{
  \inst{1} Departamento de F\'{\i}sica,
    Universidade Federal do Rio
    Grande do Norte, 59072-970
    Natal,  RN, Brazil\\
\inst{2} Departamento de F\'{i}sica, Universidade do Estado do Rio Grande do Norte, Mossor\'o--RN, Brazil\\
}

\abstract{
The present work is based on a description for the angular mometum loss rate due to magnetic braking for main-sequence stars on the relationship between stellar rotation and age. In general, this loss rate denoted by $\mathrm dJ/\mathrm dt$ depends on angular velocity $\Omega$ in the form $\mathrm dJ/\mathrm dt\propto\Omega^{q}$, where $q$ is a parameter extracted from nonextensive statistical mechanics. Already, in context of stellar rotation, this parameter is greater than unity and it is directly related to the braking index. For $q$ equal to unity, the scenario of saturation of the magnetic field is recovered, otherwise $q$ indicates an unsaturated field. This new approach have been proposed and investigated by de Freitas \& De Medeiros for unsaturated field stars. In present work, we propose a nonextensive approach for the stellar rotational evolution based on the Reiners \& Mohanthy model. In this sense, we developed a nonextensive version of Reiners \& Mohanthy torque and also compare this generalized version with the model proposed in de Freitas \& De Medeiros based on the spin-down Kawaler torque for the main-sequence stars with F and G spectral types. We use the same sample of $\sim16 000$ field stars with rotational velocity $v \sin i$ limited in age and mass. As a result, we show that the Kawaler and Reiners \& Mohanthy models exhibit strong discrepancies, mainly in relation to the domain of validity of the entropic index $q$. These discrepancies are mainly due to sensitivity on the stellar radius. Finally, our results showed that modified Kawaler prescription is compatible with a wider mass range, while the Reiners \& Mohanty model is restricted to masses less than G6 stars.}

\begin{document}

\maketitle

\section{INTRODUCTION}

It is generally accepted that the magnetic braking is the fundamental mechanism responsible for angular momentum loss through magnetic stellar winds in several classes of stars, such as main-sequence field stars and cluster stars. This break was initially suggested by Schatzman \cite{schatzman} in 1962. Later, Kraft \cite{kraft1970} showed that the behavior of mean rotation velocity of low-mass-main-sequence stars below 1.5$M_{\odot}$ is preferentially due to magnetic wind. As mentioned by Kawaler \cite{kawaler1988}, main-sequence (hereafter MS) stars with mass less than 1.5$M_{\odot}$ contain their deep surface convection zones which support magnetic field as well as acoustically controlled stellar winds.

Skumanich \cite{Skumanich}'s pioneering work reveals that the angular momentum loss of G-type
main-sequence stars for Hyades and Pleiades obeys the following relationship $\mathrm dJ/\mathrm dt \propto \Omega^{3}$, where $\Omega$ denotes angular velocity. Thus, the projected rotation velocity $v \sin i$ decreases roughly with $t^{-1/2}$, is valid only within the equatorial plane. Here, $t$ is stellar age and exponent $-1/2$ is related to exponent from $\mathrm dJ/\mathrm dt$ by simple relationship $t^{-1/2}=t^{1/(1-3)}$, where $\mathrm dJ/\mathrm dt$ is the angular momentum loss rate. In addition, unsaturated G stars are in agreement with this relationship. As quoted by Schrijver \& Zwaan \cite{sz2010}, empirical Skumanich relationship, revisited by several authors, has stood the test of time, albeit that the equations fitted to the data do differ somewhat from author to author. In contrast, Chaboyer {\it et al.}\cite{chaboyer1995} elaborated a general parametrization developed by Kawaler \cite{kawaler1988} and showed that stars that rotate substantially faster are subject to magnetic field saturation at a critical rotation rate $\omega_{sat}$. Krishnamurthi {\it et al.}\cite{kris1997} proposed the inclusion of a Rossby scaling at saturation velocity for stars more massive than 0.5$M_{\odot}$. However, Sills, Pinsonneault \& Terndrup \cite{sills2000} argue that the Rossby number is inadequate for stars with mass less than 0.6$M_{\odot}$.

Recently, Reiners \& Mothanty \cite{reiners2012} proposed a new braking approach based on the strong dependence on the stellar radius which arises from definition of the surface magnetic strength ($B_{0}\propto\Omega^{a}$). In contrast, Kawaler \cite{kawaler1988} assumes that rotation is related to the magnetic field strength ($B_{0}R^{2}\propto\Omega^{a}$) reducing dependence on the radius. According to \cite{reiners2012}, the factor of $R^{2}$ implies deep changes to the angular momentum loss law. Another important difference between the modified Kawaler and Reiners \& Mothanty torques is in the saturation threshold $\omega_{sat}$. Reiners \& Mothanty \cite{reiners2012} assume that $\omega_{sat}$ does not depend on mass and it only affects the torque in unsaturated stars. Their work showed that the angular momentum loss law for low-mass stars obeys the relationship $\mathrm dJ/\mathrm dt \propto \Omega^{5}$. For this torque, the rotational velocity decreases with stellar age according to $t^{-1/4}$.

More recently, de Freitas \& De Medeiros \cite{defreitas2013} revisited the modified Kawaler parametrization proposed by Chaboyer {\it et al.}\cite{chaboyer1995} in the light of the nonextensive statistical mechanics \cite{tsallis1988}. This new statistical theory were introduced by Tsallis \cite{tsallis1988}, and further developed by Curado \& Tsallis \cite{curado1991} and Tsallis, Mendes \& Plastino \cite{tsallis1998}, with the goal of extending the regime of applicability of Boltzmann-Gibbs (B-G) statistical mechanics that present just plainly fail when applied in out from equilibrium systems. de Freitas \& De Medeiros \cite{defreitas2013} analyzed the rotational evolution of unsaturated F- and G- field stars limited in age and mass in the solar neighborhood. They attribute to the entropic index $q$ extracted from the Tsallis formalism as a parameter that describes the level of magnetic braking as called ``braking index''. They also associated this parameter with the exponent of dynamo theory ($a$) and to magnetic field topology ($N$) through the relationship $q=1+4aN/3$. As a result, they showed that the saturated regime can be recovered in the nonextensive context assuming the limit $q\rightarrow1$. This limit is particularly important because it represents the thermodynamic equilibrium valid in Boltzmannian regime. Indeed, the torque in nonextensive version is given by $\mathrm dJ/\mathrm dt \propto \Omega^{q}$ revealing that the rotational velocity of F- and G-type main sequence stars decreases with age according to $t^{1/(1-q)}$. The values of $q$ obtained by de Freitas \& De Medeiros \cite{defreitas2013} point out that it has a strong dependence on stellar mass.

Some applications have been made to test the nonextensive approach proposed by de Freitas \& De Medeiros \cite{defreitas2013}, among them Silva {\it et al.}\cite{Silvaetal13} and de Freitas {\it et al.}\cite{defreitasetal13} using this model for open cluster stars. In Silva {\it et al.}\cite{Silvaetal13}, the authors define the index $q$ extracted from modified Kawaler parametrization as $q_{K}$, where the subscript $K$ is an abbreviation for Kawaler. In this scenario, Silva {\it et al.}\cite{Silvaetal13} suggest that the $q_{K}$ is a possible parameter that control the degree of anti-correlation between $q$ from empirical rotational distribution and the cluster age $t$ given by term $q\approx q_{0}(1-\Delta t/q_{K})$, where $\Delta t = t-t_{0}$ is the difference
between the logarithms of the cluster ages, and $q_{0}$ is the index
$q$ for the cluster with age $t_{0}$. In general, this behavior indicates that stars loss the memory of past angular momentum history when their rotational distribution becomes extensive, i.e., $q=1$. However, we wonder if this result set prevails when invoke another parametrization model of the angular momentum loss, in our specific case, the Reiners \& Mothanty torque. 

Epstein \& Pinsonneault \cite{epstein2012} compare the prescriptions for the strength of stellar wind of modified Kawaler and Reiners \& Mohanty torque with and without the disk affect for stars with mass between 0.5 and 1.0$M_{\odot}$. They conclude that the largest differences between two prescriptions appear among low mass stars. As a result, they claim that the Reiners \& Mohanty torque preserves rapid rotators of low mass stars. This effect is related to the Reiners \& Mohanty torque predicts a slower spin down than Kawaler does for the first few Myr. Epstein \& Pinsonneault \cite{epstein2012} still claim that Kawaler parametrization is indistinguishable to the Skumanich law beyond the Zero Age Main-Sequance (ZAMS). Indeed, this result is expected because the values adopted for $a$ and $N$ imply that angular velocity decays with time asymptotically as $\Omega\sim t^{-1/2}$. In contrast, for the Reiners \& Mohanty torque using same values of $a=1.5$ and $N=2$, the spin down rate decreases to a weaker power law, i.e., $\Omega\sim t^{-1/4}$. In fact, the Kawaler (standard) framework of angular momentum loss rate described above is limited to stars below late-F stars. For stars above early-F a simple scaling of the solar torque by $\omega^{3}$ does not explain the fast rotators in young stars, in this case, saturation in the angular momentum loss is necessary. In both loss laws, rotation rate from saturated stars diminishes exponentially rapidly to critical saturation velocity. In general, according to de Freitas \& De Medeiros \cite{defreitas2013}, as the exponent of $\omega$ is sensitive to stellar mass, we obtain fluctuations this exponent in narrower mass range than those above defined. 

In this study we compare the Kawaler and Reiners and Mothanty torques in the context of the Tsallis'
nonextensive statistical mechanics. In Section 2, we revisit and generalize the parametric Reiners and Mothanty model for angular
momentum loss by magnetic stellar wind, as well as, we compares both models. Finally, conclusions are presented in Section 5.

\section{Generalized Reiners \& Mohanty torque}
Reiners \& Mothanty \cite{reiners2012} propose a new approach for the stellar magnetic braking scenario. This approach considers two major deviations from modified Kawaler prescription (see \cite{defreitas2013} for a review). As mentioned by Epstein \& Pinsonneault \cite{epstein2012}, first Reiners \& Mothanty\cite{reiners2012} argue that rotation is related to the magnetic field strength rather than the surface magnetic flux as adopted in Kawaler \cite{kawaler1988}. This dependence in $R^{2}$ modifies the structure of the angular momentum loss law, indicating that the Kawaler model presents a weak dependence on the stellar radius. And second, these authors assume that the saturation angular velocity $\omega_{sat}$ does not depend on stellar mass. In this case, $\omega_{sat}$ is insensitive to structural differences in stellar interior. In this context, Reiners \& Mothanty \cite{reiners2012} restrict the rate of angular momentum loss to radial fields and a dynamo law $a=1.5$ for unsaturated regime, given by following equations
\begin{eqnarray}
\label{loss2}
\frac{\mathrm dJ}{\mathrm
dt}=-C\left[\Omega\left(\frac{R^{16}}{M^2}\right)^{1/3}\right]
\end{eqnarray}
for $\Omega\geq\omega_{sat}$. Already, for $\Omega<\omega_{sat}$, we have

\begin{eqnarray}
\label{loss1}
\frac{\mathrm dJ}{\mathrm
dt}=-C\left[\left(\frac{\Omega}{\omega_{sat}}\right)^{4}\Omega\left(\frac{R^{16}}{M^2}\right)^{1/3}\right]
\end{eqnarray}
with $C=\frac{2}{3}\left(\frac{B^{8}_{sat}}{G^{2}K^{4}_{V}\dot{M}}\right)^{1/3}$.

Here, $G$ is the gravitational constant, $\omega_{sat}$ is the threshold angular velocity beyond which
saturation occurs, $B_{sat}$ denotes the saturation field strength, $K_{V}$ is a calibration constant and $R$, $M$ and
$\dot{M}$ represent the stellar radius, mass and mass-loss rate in units
of $10^{-14}M_{\odot}$yr$^{-1}$, respectively. 

\begin{figure*}
\begin{center}
\includegraphics[width=0.65\textwidth]{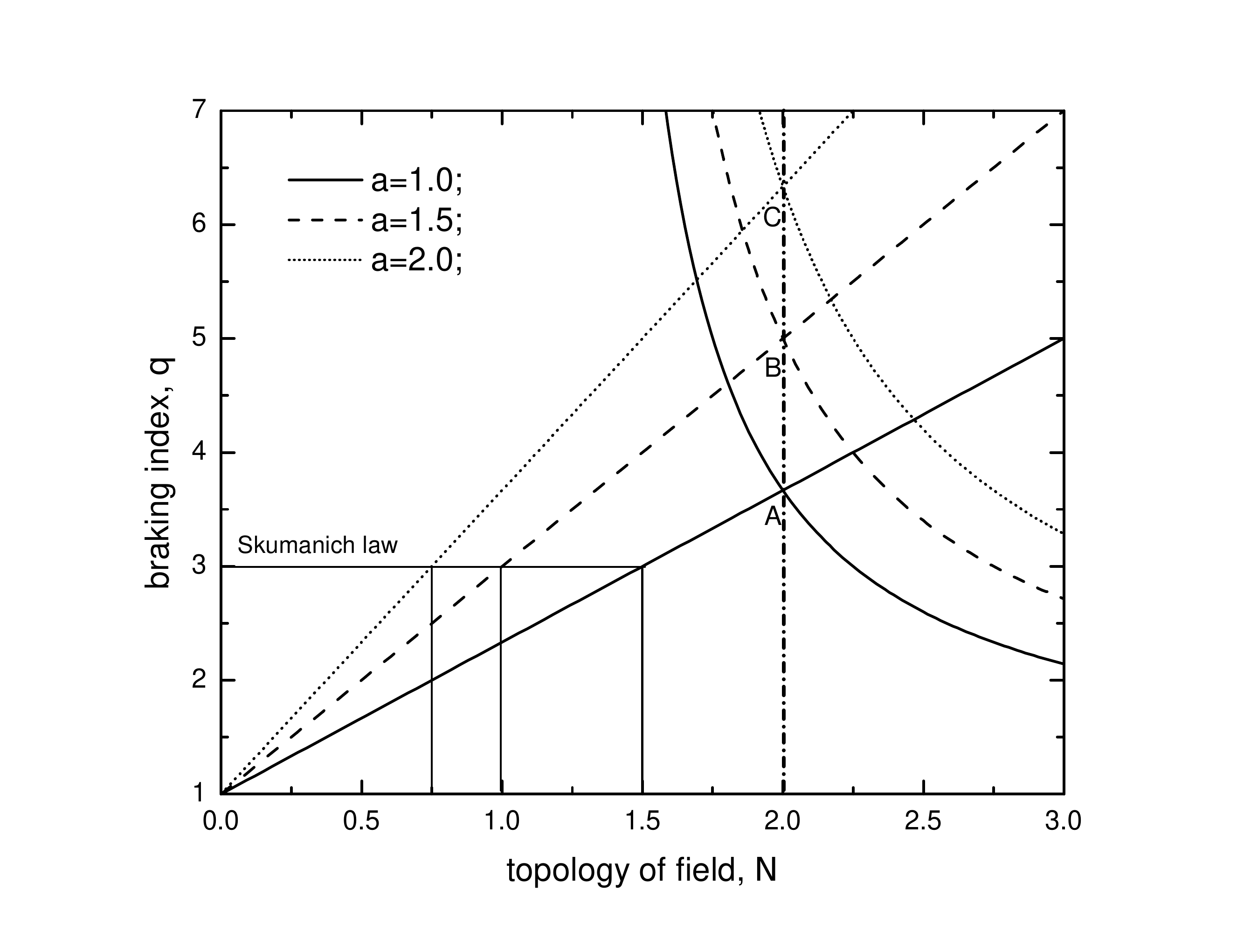}
\end{center}
\caption{Plot $q$ from model against the field topology $N$. Straight lines represent the Kawaler entropic index, while curves denote Reiners \& Mothanty one. The vertical dashed dot line shows that exist only one valor for $N$ where $q$'s are equals, i.e., when $N=2$. The letter $A$ represents the lowest value of $q_{RM}$. Already, $B$ denotes the Reiners \& Mothanty \cite{reiners2012} exponent, i.e., $t^{-1/4}$. And $C$ is the maximum possible value of $q$ among the models. Plot also shows that the Skumanich law can not be represented on the generalized Reiners \& Mothanty torque.
}
\label{fig1}
\end{figure*}

For a future comparison with the parameters obtained in nonextensive Kawaler model presents in de Freitas \& De Medeiros \cite{defreitas2013}, we eliminate the restrictions in $a$ and in the magnetic field geometry $N$ adopted by Reiners \& Mothanty \cite{reiners2012}. After a rather tedious algebraic manipulation, the generalized Reiners \& Mothanty torque, for the different saturation regimes, where the parameters $a$ and $N$ are free, is given by
\begin{eqnarray}
\label{loss3}
\frac{\mathrm dJ}{\mathrm
dt}=-C^{'}\left[\Omega\left(\frac{R^{16-8N}}{M^2}\right)^{1/(3-4N)}\right]
\end{eqnarray}
for $\Omega\geq\omega_{sat}$. In this regime, equation above is not depend on the parameter $a$. Already, for $\Omega<\omega_{sat}$, we have

\begin{eqnarray}
\label{loss4}
\frac{\mathrm dJ}{\mathrm
dt}=-C^{'}\left[\left(\frac{\Omega}{\omega_{sat}}\right)^{8a/(3-4N)}\Omega\left(\frac{R^{16-8N}}{M^2}\right)^{1/(3-4N)}\right]
\end{eqnarray}
with $C^{'}=\left[\left(\frac{2}{3}\right)^{4N}C^{3}\right]^{1/(3-4N)}$. If $a=1.5$ and $N=0$ (later we will replace $N = 2$), we recovered the Reiners \& Mohanty torque.
 
In the present study, we assume that since the moment of inertia $I$ and
stellar radius $R$ changes slowly during main-sequence evolution, the angular
momentum loss law can be simplified by considering the limit of $\mathrm
dJ/\mathrm dt$ for constants $I$ and $R$, that is, this loss law is fully specified by the rotational velocity, while the star is braked by the stellar wind \cite{bouvier1997}. We also consider that in the
absence of angular momentum loss, equatorial rotational velocity $v$ can
be determined by the simple condition of angular momentum conservation,
denoted by
\begin{equation}
\label{loss5}
J=\left(\frac{I}{R}\right)_{const.}v,
\end{equation}
furthermore, we assume that stars spin down as a solid body.

Two pairs of equations were combined to obtain the time dependence of $v$.
First, from eqs. (\ref{loss3}) and the derivative of (\ref{loss5}), we have
\begin{equation}
\label{loss6}
v(t)=v_{0}\exp \left[-\frac{(t-t_{0})}{\tau_{1}}\right], \quad (t_{0}\leq t<
t_{sat})
\end{equation}
with
\begin{equation}
\label{loss6x}
\tau_{1}\equiv\left[\frac{C^{'}}{I}\left(\frac{R^{16-8N}}{M^{2}}\right)^{1/(3-4N)}\right]^{-1}
\end{equation}
for the saturated domain.

On the other hand, by combining eqs. (\ref{loss4}) and the derivative of (\ref{loss5}), we
obtain the equation
\begin{eqnarray}
\label{loss8}
v(t)=v_{sat}\left[1+\frac{(t-t_{sat})}{\tau}\right]^{-(3-4N)/8a},
\quad (t\geq t_{sat})
\end{eqnarray}
with
\begin{equation}
\label{loss4xy}
\tau\equiv\left[\frac{8a}{3-4N}\frac{C^{'}}{I}\left(\frac{R^{16-8N}}{M^{2}}\right)^{1/(3-4N)}\right]^{-1}.
\end{equation}

In this respect, the $\tau_{1}$ and $\tau$ are the MS spin-down
timescales in the saturated and unsaturated regimes, respectively. The
other parameters, $t_{0}$, $v_{0}$ and $t_{sat}$, are the age at which the
star arrives on the MS, the rotational velocity at that time and
subsequent age at which the star slows down to values below the critical
velocity $v_{sat}$, respectively \cite{reiners2012}. Time $t_{sat}$ can
be determined by setting $v(t)=v_{sat}$ in eq. (\ref{loss6}). Thus, we
have
\begin{equation}
\label{tcrit}
t_{sat}=t_{0}+\tau_{1}\ln\left(\frac{v_{0}}{v_{sat}}\right).
\end{equation}

Equation (\ref{loss6}) reveals that in the saturated regime the star slows down
exponentially from time $t_{0}$ up to velocity $v_{sat}$ in time
$t_{sat}$. Already, eq. (\ref{loss8}) shows that from this time onward, the
spin-down rate decreases to a power law given by $-(3-4N)/8a$. Thus, the star
remains within a factor of a few $v_{sat}$ for the remainder of its MS
lifetime.

\section{Nonextensive approach for stellar rotation-age relation for the unsaturated regime}

Applying Tsallis statistics \cite{tsallis2004}, we can assume that eq.
(\ref{loss6}) follows a simple linear differential equation in the form
\begin{equation}
\label{loss78}
\frac{\mathrm d}{\mathrm dt}\left(\frac{v}{v_{0}}\right)=-\lambda_{1}\left(\frac{v}{v_{0}}\right),
\end{equation}
from which the solution is given by
\begin{equation}
\label{rela4}
v(t)=v_{0}\exp\left[-\lambda_{1}(t-t_{0})\right].
\end{equation}

According to Lyra {\it et al.}\cite{lyra}, in contrast to the exponential behavior
presented in eq. (\ref{rela4}), eq. (\ref{loss78}) must be replaced with a
slower power law at critical points where long correlations develop. In
this context, a similar expression can be created to characterize the
possible behavior of stellar rotation distribution in an unsaturated
regime. As proposed by de Freitas \& De Medeiros \cite{defreitas2013}, the spin down law of a group of the stars limited by mass is usually given in the generalized form
\begin{equation}
\label{loss8x}
\frac{\mathrm d}{\mathrm
dt}\left(\frac{v}{v_{sat}}\right)=-\lambda_{q}\left(\frac{v}{v_{sat}}\right)^{q}
\quad\ (\lambda_{q}\geq 0; q\geq 1),
\end{equation}
if $q=3$ the spin down is determined by torque generated by magnetically controlled stellar winds.

Integrating the equation above, we have
\begin{equation}
\label{loss8a}
v(t)=v_{sat}\left[1+(q-1)\lambda_{q}(t-t_{sat})\right]^{\frac{1}{1-q}},
\end{equation}
where $\left[1+(1-q)x\right]^{1/(1-q)}\equiv \exp_{q}(x)$ is known in nonextensive scenario as $q$-exponential (see \cite{borges2004} for a review of $q$-calculus). 
In our context, $\lambda_{q}$ indicates the braking strength.

Thus, in the nonextensive scenario, eq. (\ref{loss8}) is well described as
a non-linear differential equation in form (\ref{loss8x}) with solution
(\ref{loss8a}). When $q=1$, eq. (\ref{rela4}) is recovered, indicating
that a system governed by a saturated regime is in thermodynamic
equilibrium. By contrast, when $q$ differs from unity, a system controlled
by unsaturation is out of equilibrium.

In agreement with our unsaturated regime model and replacing $N$ by $N-2$, we find that the ``braking index'' $q_{RM}$ can
be described by a pair ($a,N$) given by
\begin{equation}
\label{loss10}
q_{RM}=1+\frac{8a}{4N-5},
\end{equation}
whereas for saturated regime, $q_{RM}=1$ due to the usual exponential function. As showed in table \ref{tab1}, the parameters $a$ and $N$ are as a function of spectral type, i.e., the braking index $q$ 
is related to stellar mass and evolution of stellar magnetic field. In Fig. \ref{fig1}, we have the plot of index $q$ against the field topology $N$. Straight lines represent the Kawaler entropic index given by identity $q_{K}\equiv 1+4aN/3$, while that the curves denote the Reiners \& Mothanty prescription given by term (\ref{loss10}). For such model, figure shows lines and curves as an iso-$a$ (specifically $a=1$, $a=1.5$ and $a=2$). The vertical dashed dot line shows that exist only one value for $N$ where $q_{K}$ and $q_{RM}$ are equals, i.e., $N=2$. From referred Figure, the letter $A$ represents the possible lowest value of $q_{RM}$. Already, $B$ denotes the exponent adopted by Reiners \& Mothanty \cite{reiners2012}, i.e., $q_{RM}=5$. And $C$ is the maximum possible value of $q$ among the models. Figure also shows that the Skumanich law can not be represented on the generalized Reiners \& Mothanty torque.

In general, the $q$-parameter is related to the degree of nonextensivity that emerges
within thermodynamic formalism proposed in Tsallis \cite{tsallis1988}. As
revealed in eq. (\ref{loss10}), $q_{RM}$ is a function of magnetic field
topology $N$ and dynamo law $a$, which depend on stellar evolution.
According to Kawaler \cite{kawaler1988} and de Freitas \& De Medeiros \cite{defreitas2013}, small $N$ values result in a weak wind
that acts on the MS phase of evolution, while winds with large $N$ values
remove significant amounts of angular momentum early in the
Pre-Main-Sequence (PMS). In this phase, for a given mass, maximum
rotational velocity $v$ decreases as $N$ increases. This result is
significant because, within a thermostatistical framework in which eq.
(\ref{loss8}) naturally emerges and for a given value of $a$ and $N$, we obtain the
scale laws found in the literature, such as those proposed by
Skumanich \cite{Skumanich}, Soderblom {\it et al.}\cite{Soderblom1991} and Reiners \& Mothanty \cite{reiners2012}.

\begin{table}
\scalefont{0.8}
\caption{Best parameter values of our unsaturated model using
eq. (\ref{loss8a}). The values of $N$ using eq. (\ref{loss10}) also are shown.}
\label{tab1}
\renewcommand{\arraystretch}{1.5}
\begin{tabular}{ccccc}
\hline \hline
Stars & $\left\langle M/M_{\odot}\right\rangle$ & $q_{RM}$ & $N_{a=2};N_{a=1}$ & $a_{N=0};a_{N=2}$\\
\hline \hline
F0-F5 & 1.36 & 1.80$\pm$0.1 & $6.25 ; 3.75$ & $-0.5 ; 0.3$\\
F6-F9 & 1.22 & 1.96$\pm$0.2 & $5.42 ; 3.33$ & $-0.6 ; 0.36$\\
G0-G5 & 1.11 & 2.18$\pm$0.2 & $4.64 ; 2.95$ & $-0.74 ; 0.44$\\
G6-G9 & 0.98 & 4.38$\pm$0.6 & $2.43 ; 1.84$ & $-2.11 ; 1.26$\\
\hline
All F &     &   1.91$\pm$0.1 & $5.65 ; 3.45$ & $-0.57 ; 0.34$\\
All G &     &  2.18$\pm$0.4  & $4.64 ; 2.95$ & $-0.74 ; 0.44$\\
\hline
\end{tabular}
\end{table}

\section{CONCLUSIONS}\label{conclusions}
In summary, our study presents a new approach on the angular momentum loss law found in literature. The present analysis shows that the generalized torques of Kawaler and Reiners \& Mothanty are clearly antagonistic, except for $N=2$. Table \ref{tab1} demonstrates clearly that the modified Reiners \& Mothanty model is valid only for spectral types beyond G6. According to Fig. \ref{fig1}, the lowest valor of $q_{RM}$ is 3.67. As a result from this figure, the Skumanich law can not be represented on the generalized Reiners \& Mothanty relationship, i.e., this model does not explain the rotational behavior of solar type stars. This limitation is mainly due to sensitivity on the stellar radius given by term $R^{16/3}$, i.e., the angular momentum loss rate would have become entirely dependent on the stellar radius. In Kawaler model, this dependence is given by maximum term $R^{2}$, where $\frac{\mathrm dJ}{\mathrm
dt}$ depends entirely on the stellar velocity for unsaturated stars. However, our results reveal which Reiners \& Mothanty model is in good agreement for low-mass stars, specifically, M-type stars. In other words, the Reiners \& Mothanty model is a good approximation for fully convective stars.

As mentioned by de Freitas \& De Medeiros \cite{defreitas2013}, the entropic index $q$ is an indicator of the memory of system or, in our case, the memory of primordial angular momentum. As $q$ increases from F to M, very low-mass stars maintain this type of memory. In fact, the memory grows at same scale of $q$. In contrast, because the much shorter interaction of the young star with its circumstellar disc, higher mass stars do not maintain the memory of initial angular momentum for long time.  According to de Freitas \& De Medeiros \cite{defreitas2013} and our present results, both nonextensive models show that the Kawaler framework is compatible with a wider mass range, while the Reiners \& Mothanty model is restricted to low-mass stars less than G6 masses. This result suggests that the decrease in mass would lead to weaker power law than the Skumanich law. Finally, the results point out that a strong dependence of angular momentum loss rate on stellar radius is limited rather to very low-mass stars encountering discrepancies when applied to F and G stars.  

\acknowledgments
Research activity of the Stellar Board of the Federal University of Rio Grande do Norte (UFRN) is supported by continuous grants from CNPq and FAPERN Brazilian agency. B.B. Soares and J.R.P. Silva acknowledge financial support from the Programa Institutos Nacionais de Ci\^encia e Tecnologia (MCT-CNPq- Edital no. 015/2008).

\end{document}